\documentclass[12pt,preprint]{aastex}

\slugcomment{Submitted to Ap.J.}

\shorttitle{Field Ellipticals}
\shortauthors{Smith, R.M. et al.}

\begin{document}

\title{A Sample of Field Ellipticals}

\author{Rodney M. Smith}
\affil{Dept. of Physics and Astronomy, University of Wales, Cardiff, PO Box 913,
Cardiff, CF24 3YB, U.K.}
\email{R.Smith@astro.cf.ac.uk}

\author{Vicent J. Mart\'{\i}nez}
\affil{Observatori Astron\`omic, Universitat de Val\`encia, 
Apartat de Correus 22085, E-46071 Val\`encia, Spain}
\email{Vicent.Martinez@uv.es}

\and

\author{Matthew J. Graham}
\affil{Astronomy Department, 1200 E California Blvd, California Institute of
Technology, Pasadena, California 91125, U.S.A.}
\email{mjg@cacr.caltech.edu}

\begin{abstract}
Using well-defined selection criteria applied to the LEDA galaxy catalogue
we have constructed a sample of elliptical galaxies that can be taken to lie
in the field. Such criteria can easily be applied to theoretical
simulations for direct comparison with observations. 
The variation of the number of `isolated' ellipticals with 
selection criteria is also investigated. A preliminary study of the
environment of the field ellipticals shows that, in the mean, they
are surrounded by a population of dwarf galaxies,
out to projected radii of at least 500 kpc, with a radial density profile
of $r^{-0.6\pm 0.2}$ and a luminosity function slope of $\alpha \sim 
-1.8$. 
The results are compared and contrasted to the satellite population around
isolated spiral galaxies.
\end{abstract}


\section{Introduction}

From the many observations of the two-point correlation function, both in its
angular (e.g. Lidman and Peterson 1996, Maddox et al. 1990) and spatial 
(e.g. Ratcliffe et al. 1998, Giuricin et al. 2001) form, it is now 
well established that galaxies are clustered out to distances of at
least 10 Mpc. In addition, numerous studies of
the distribution of bright galaxies (e.g. Dressler 1980) show that elliptical
galaxies preferentially occur in regions of high galactic density
whilst
spiral galaxies dominate in the field. These basic observations
have led to a multitude of theoretical simulations to
explain the observed clustering properties and morphological
segregation
(e.g. Baugh et al. 1996). A major implication of the majority of these 
models is that elliptical galaxies are formed from the merger of many
sub-clumps during the early stages
of the evolution of the Universe. Although not ruling out the
presence of elliptical galaxies in low density environments, the
hierarchical models
suggest that on average they are very different from the cluster ellipticals,
with likely evidence of recent star-formation and/or merger events.
There have been many studies of elliptical galaxies in low-density
environments with somewhat inconclusive results, with some studies
suggesting only minor star formation at low redshifts (e.g. Silva and
Bothun 1998, Bernardi et al. 1998) whilst other studies have shown
strong
evidence for recent merger/star formation activity (e.g. Treu et al.,
1999, 2001, Kuntschner et al. 2002).

A major problem with the current comparisons between theory and
observation is the lack of a consistent
definition of a field galaxy. Several of the studies (e.g. 
Treu et al. 1999, 2001,
2002, Aars et al. 2001) use redshift surveys of the brighter galaxies to
derive a sample of isolated ellipticals. However, the incompleteness of 
the redshift catalogues may lead to the inclusion of several ellipticals
that have close neighbours. This has led many to a final visual
inspection to confirm their isolated nature, destroying the objectiveness
of the selection criteria. Only using an extensive redshift survey can an
objective sample be constructed (e.g. Kuntschner et al. 2002) but even
then incompleteness in the catalogue can lead to erroneous selection
of non-field galaxies.  The studies have also concentrated on the 
properties of the galaxies themselves, with little, if any, analysis
of the local environment of the elliptical which, from the
morphology-density relationship, is likely to also have a very
significant effect on the properties of the elliptical galaxy.

In this paper, we use an objective definition of a field 
galaxy applied to an all-sky galaxy catalogue that can also
be applied to the theoretical simulations. We also investigate the
variation of the number of field ellipticals with selection criteria.
Using all-sky photographic surveys we have made 
a preliminary study of the environment of these galaxies in a 
search for a surrounding faint dwarf galaxy population.

\section{Determination of the Field}

The presence of satellite galaxies around much brighter galaxies and the
possible presence of large numbers of dwarf galaxies in the field greatly 
complicates the selection of a field sample.
At present, our knowledge of the faint end of the galaxy luminosity function (LF)
is very uncertain. Studies of clusters have produced widely varying
results, although evidence is increasing of a relationship between the
local galaxy density and the gradient of the function at the faint end
(e.g. Driver et al. 1998).
The investigations suggest that in low density regions on the
outskirts of clusters there should be a large
population of dwarf galaxies. This may imply that in the field a large
dwarf population also exists. Results from the recent major redshift surveys
(e.g. 2dF, Norberg et al. 2002) suggest that the field LF is relatively flat.
However, incompleteness and a selection bias against low
surface brightness galaxies may lead to a flatter slope.
Using the sample of galaxies selected by Zaritsky et al. (1993, 1997; hereafter ZSFW)
Morgan et al. (1998) found evidence that, on average, 
the faint end slope of the LF in the fields of isolated spirals is steep,
with $\alpha \sim -1.8$.
 Although this is consistent with that found in the
outer regions of clusters, there is a major discrepancy between this
result and our knowledge of the LF in the field and the
Local Group, the latter having
a flat faint-end of the LF (Mateo 1998). Roberts et al. (2003)
have also found evidence that the slope of the field LF is flat.
With the uncertainty in the field
LF beyond $M_B \sim -17$ (for $H_0 = 70$ km s$^{-1}$ Mpc$^{-1}$ as adopted
thoughout) and the general consistency of the LF shape
at the bright end (e.g. Driver and de Propris 2003), we concentrate here 
on defining a field galaxy as 
that which does not have nearby bright ($M_B \sim -17$) neighbours
such that it has not been seriously disturbed by its current local
environment.

ZSFW, in their
study of isolated spiral galaxies, used two criteria to select their sample.
To be isolated, the magnitude difference between a neighbour and the `parent'
must be greater than 0.7 mag for galaxies within a projected distance of 1 Mpc
or greater than 2.2 mag within 500 kpc.
With an average separation of galaxies of about 1.5 Mpc (e.g. Nolthenius
and White 1987), then
the first of these criteria should ensure that the haloes do not interfere
whilst the second is the true isolation criteria, ensuring that the galaxy
does not lie in a cluster or a rich group. The application of these criteria is 
shown as a schematic in Fig. 1. Using these criteria the Magellanic
Clouds would count as satellites but the Local Group would fail the criteria
due to the close spacing of M31 and the Milky Way. Even the rather isolated elliptical
galaxy NGC 720 (Dressler et al. 1986) would not be considered 
a field galaxy under our demanding criterion.
ZSFW results, with a lack of any neighbouring satellites beyond 
a galactocentric distance of about 500 kpc,
lent strong support to this classification of isolated galaxies.

\section{The Sample}

The availability of catalogues of large numbers of galaxies now makes it 
possible to investigate the dependency of galaxy morphology on
environmental properties. Primary sources of information on galaxies are
the Lyon Extragalactic Database (LEDA) and the
NASA/IPAC Extragalactic Database (NED). In this study we use the LEDA catalogue.
This database, constructed from a collection of
sub-catalogues and other sources, currently contains data for over one million
galaxies. More recent catalogues, such as that from the APM and the SDSS,
concentrate primarily on the fainter galaxies and therefore are not complete
at the bright end and cannot be reliably used for the selection of 
bright, isolated, galaxies over the whole sky.
Using the LEDA
catalogue we have derived samples of elliptical galaxies satisfying
several criteria. Firstly, the redshift of the primary galaxy must be less than
$10,000$ km s$^{-1}$ to ensure that the sample is approximately complete.
To ensure that a reasonably accurate value for
the distance could be derived assuming a uniform 
Hubble flow, an inner redshift 
limit of $1500$ km s$^{-1}$ was also applied. 
Secondly, the absolute magnitude of the galaxy must be less than $
M_B \le -19$. This criterion was applied for several reasons. 
Galaxies brighter than this can be taken to be `normal' and applying the 2.2
magnitudes fainter criteria later ensures that the faint end of the LF, with 
all its uncertainties, is not reached for the neighbours. Again it also 
ensures, with the redshift data, that the catalogue is 
approximately complete. Other criteria
applied are that the selected galaxies lie above a galactic latitude of 
$|25^\circ|$ to minimise the effects of Galactic absorption and
to ensure that only elliptical galaxies are selected we applied a $t<-4$
type criterion to the LEDA database. Misclassification of galaxies is
always a problem, particularly for elliptical galaxies where the
presence of a disk or dust may lead to an erroneous morphological
assignment, and it is possible that some S0's may be
included in this sample but the percentage should be very small. A visual
inspection of the galaxies, together with a literature search, was 
undertaken to verify the morphological classification of the galaxies as
ellipticals and also minimise the contamination. As the morphology of
some of the galaxies was uncertain they have been retained in the catalogue
and await a more detailed imaging study for an accurate classification.
The heterogeneous nature of the construction of the LEDA catalogue almost 
certainly leads to an uncertain incompleteness limit. As the limiting apparent
magnitude of our sample at the selected redshift cutoff is 16.8, it is highly 
likely that some galaxies will be missed
due to a lack of redshift information. Although precluding a detailed
statistical study of the population this does not, however, detract
from our original desire to obtain a sample of elliptical galaxies in low
density regions.

The total number of elliptical
galaxies in the LEDA database that satisfy the selection criteria
described above is 940.
For each of these galaxies we have searched the LEDA database for galaxies that
satisfy the ZSFW selection criteria. The first of these is that any galaxy 
within 1 Mpc should be at least 8 times (=2.2 mag) fainter than the primary. 
Secondly, the LEDA database was searched for any galaxy within a specified
projected distance that was less than twice as faint as the primary candidate.
Unlike previous studies, no redshift information was included in the selection
process and thus these distances are projected.
This is a much stricter criterion than many other studies, but 
ensures that any galaxies that satisfy the constraints are truly isolated.
The 32 elliptical galaxies that we determine to be isolated are listed in Table 1,
together with notes on the individual galaxies, taken from the NED database. 
A visual inspection of the Digitized Sky Survey scans was also undertaken to
ensure that there were no bright galaxies 
in the field that had been missed in the LEDA
catalogue.

\section{Changing the Parameters}

How does the percentage of isolated galaxies vary as a function of 
magnitude limit of primary? or radial limits? 
To estimate the dependency of the field galaxy sample size with selection 
criteria we have applied an identical technique to that described above but have
varied the inner radius cut-off from the fixed 500 kpc limit 
of ZSFW. 
The variation of the percentage of the morphological
sample that are classified as `field'  with the inner radius cut-off is 
plotted in Fig. 2.
Although closely related to the two-point correlation function, the addition
of magnitude and redshift information in the production of this plot
precludes a direct comparison.
It is clear that there are elliptical galaxies that do not have companions
within 2.2 magnitudes out to at least 1 Mpc.

To investigate the effect of incompleteness in the LEDA catalogue at fainter
magnitudes,
we have also used the same isolation and selection criteria except using
an absolute magnitude of $M_B < -20.5$, corresponding to an apparent
magnitude limit of $B=15.3$ at a redshift of $10000$ km s$^{-1}$. 
The variation
in the percentage of galaxies classified as isolated is shown
in Fig. 2 as the dotted line. Although the number of ellipticals in this
brighter sample is much smaller, 423 compared to 940, 
the percentage of galaxies that satisfy the
isolation criteria is significantly higher at all radii than the fainter 
sample. There are several possible reasons for this. Firstly, 
by going to brighter magnitudes for the parent then, applying a magnitude limit
to the companions, we sample only the brighter part of the galaxy 
number counts. 
Thus, a brighter sample will have many fewer
companions than a fainter one, thus increasing the likelihood of the
galaxy to be classified as isolated. Secondly, it 
is possible this is due to a selection effect in the catalogue
construction. Finally,
it may be an inherent luminosity segregation in the formation process of 
elliptical galaxies, similar to the creation of an anomalously
bright galaxy in the core of some clusters, as seen in cD-dominated or 
Bautz-Morgan Type I clusters.
At present we cannot distinguish between these three possibilities. However,
a similar investigation for spiral galaxies shows that by going to the 
brighter magnitude the percentage classified as field increases by a 
similar amount as for the elliptical sample. It is not, therefore, an effect
that is due to an inherent property of the morphology of the galaxy.

\section{The Environment of Field Ellipticals}

Although the galaxies in Table 1 are isolated from other bright galaxies it
is possible that they are surrounded by a halo of fainter, dwarf, galaxies.
As a preliminary investigation of the environment of these galaxies we have
used the technique first employed by Holmberg (1969) and extended by
Phillipps and Shanks (1987) and
Lorrimer et al. (1994). Due to their clustering properties, any excess in the
number of galaxies seen in the field of the parent should be due
primarily to objects of a similar redshift. This technique has been used
extensively to determine the slope of the faint end of the LF 
in rich clusters (e.g. Driver et al. 1998) although inherent
variations in the background may lead to an erroneous steepening of the
slope (Valotto et al. 2001). However, this effect should not be important
for individual galaxies although in this case the expected low
numbers of companions will lead to large statistical errors. By stacking
a number of galaxies, as used by ZSFW and Morgan et al. (1998), 
the statistical errors can be
reduced and a `mean' profile obtained.

Using the data publicly available from the APM plate-scanning machine
we have detected all galaxies within  a projected distance of 500 kpc of the
parent. An absolute magnitude limit of $M_{Bj}=-14.6$ for the surrounding
galaxy population (assuming they are all at the redshift of the parent) and
a redshift limit of $6500$ km s$^{-1}$ for the primary was applied
to ensure that the APM scans were reasonably complete for high
surface-brightness objects. It is well-known that at low surface-brightnesses
the catalogue is incomplete. At magnitudes brighter than
$M_B=-16.8$ the criteria used to select the parent sample will
lead to an imcompleteness in the sample of dwarfs. A
total of 10 galaxies in the sample of 32 had APM data suitable for this study.
The resultant `mean' radial density profile of the dwarfs surrounding the 
sample of 10 parents 
is shown in Fig. 3. It is clear that there is a significant excess of dwarf
galaxies out to at least 500 kpc.
To obtain an estimate of the dwarf population requires accurate
subtraction of the
contaminating background population. Incorrect subtraction can lead to
widely varying values of the LF slope. With ZFSW finding very few satellites
beyond 500 kpc we use the outer values of the radial profile as an estimate of
the background. Fitting a power law to the
resulting background-subtracted galaxy counts gives a power-law
slope of $-0.6 \pm 0.2$.
This is similar to the slope found for late-type galaxy satellites by
Lorrimer et al. (1994)
but less steep than that found by them for early-type galaxies. However, they
found a weak dependence of the slope on the satellite luminosity, with fainter
galaxies having a flatter slope. Extrapolating their
results to the magnitude limits reached in this study,
there is good agreement with the
value presented here. They did not find such a 
luminosity dependence of the slope for late-type
galaxies.

There are, in total, an average of
$45\pm 15$ dwarfs within 500 kpc of each primary down to the limiting
magnitude of 14.6 and $19 \pm 6$ with $-16 < M_B < -15$.
Brighter than $-16$ the number of satellites agrees
with the values of Lorrimer et al (1994). 
Comparing the number of faint dwarfs to the
values for brighter satellites implies a steep luminosity function ($\alpha
\sim -1.8$), in approximate
agreement with that found for poor clusters (e.g. Driver et al. 1998) and also
the value derived by Morgan et al. (1998) for isolated spirals.  
This lends some support to the CDM model of
hierarchical structure formation (e.g. White and Frenk 1991), where there
should be an abundance of small dark matter halos. However, the field
luminosity function has a slope of $-1.2$ (e.g. Norberg et al. 2002, Davies et al.
2003) suggesting
the dominance of dwarf galaxies varies considerably between environments.
The errors in Fig. 3 are much larger than Poissonian, suggesting that there
are intrinsic variations in the dwarf population surrounding the parent
ellipticals. However, the sample size of dwarfs around individual
ellipticals is too small for any significant conclusions to be drawn.
If the slope of the LF is steep around field ellipticals, a deeper and
higher resolution imaging study should enable the unambiguous detection of this
dwarf population and possibly investigate the galaxy-to-galaxy variation of the
dwarf density. A dynamical study, as undertaken by ZSFW, is also required
to determine which of the brighter neighbours are  truly
satellites of the central elliptical and hence obtain an estimate of the size
(and mass) of its halo.

\section{Conclusions}

We have identified a sample of elliptical galaxies that lie in
regions where the local density of bright ($M_B < -17$) galaxies
is very low, indicating that such objects are not exclusively associated with
groups or clusters of galaxies. A study of the local environment around them
shows an excess of faint galaxies, presumably satellites, out to a 
projected distance of at least 500 kpc and with a projected density varying
as $r^{-0.6\pm 0.2}$. 
The numbers of these dwarfs suggests a steep faint end of the luminosity
function, in contradiction to that found for the field but in good
agreement with that found for the outer regions of clusters.
A considerable number of questions remain which can only be answered through a more
detailed study of these objects and their environment.

\acknowledgments{
We want to thank Prof. Carlos Frenk for embarking
us in the analysis of field elliptical galaxies, their environments and
their satellites. R. Smith wants to thank 
the Valencia University Astronomical Observatory for its hospitality
during a visit in which part of the work was done.
This work has been supported by the Spanish MCyT 
projects AYA2000-2045 and AYA2003-08739-C02-01 and by the Generalitat Valenciana
ACyT project CTIDIB/2002/257. 

We have made use of the LEDA database
(http://leda.univ-lyon1.fr)}

\clearpage
\begin{table*}
\begin{center}
\caption{Isolated ellipticals\label{tbl-2}}
\begin{tabular}{rlcccccc}
\tableline\tableline
&&&&&&&\\
&&&&&&&\\
PGC & Other Name & RA(2000) & Dec(2000) &Vel (km s$^{-1}$)& B & Notes & \\
&&&&&&&\\
&&&&&&&\\
8160 & NGC 821 & 02 08 21.1 & +10 59 42 & 1735 & 11.90 & Possible stellar disk &\\
37366 & NGC 3962 & 11 54 40.1 & $-13$ 58 30 & 1815 & 11.75 & Gaseous disk & \\
63620 & IC 4889 & 19 45 15.8 & $-54$ 20 37 & 2574 & 12.02 & Dust disk - possible S0&\\
71730 & IC 5328 & 23 33 17.0 & $-45$ 01 01 & 3137 & 12.29 & Group member&\\
60536 & NGC 6411 & 17 35 32.4 & $+60$ 48 48 & 3806 & 13.02 & &\\
27600 & NGC 2954 & 09 40 24.1 & +14 55 22 & 3821 & 13.60 & & \\
72867 & NGC 7785 & 23 55 19.0 & +05 54 57 & 3808 & 12.62 & Extended envelope & \\
11274 & NGC 1162 & 02 58 56.0 & $-12$ 23 55 & 3939 & 13.49 & & \\
15406 & NGC 1600 & 04 31 39.8 & $-05$ 05 10 & 4688 & 12.04 & Group member &\\
62342 & NGC 6653 & 18 44 38.4 & $-73$ 15 48 & 5163 & 13.35 & Possible SA(rs)0 &\\
1037 & NGC 57 & 00 15 30.9 & +17 19 42 & 5440 & 12.87 & & \\
7252 & NGC 741 & 01 56 21.0 & +05 37 44 & 5561 & 12.30 & Group member &  \\
28220 & NGC 3017 & 09 49 03.0 & $-02$ 49 19 & 6229 & 14.45 & & \\
9858 & IC 1819 & 02 35 41.8 & +04 03 06 & 6393 & 15.29 &  Probable S0 & \\
45976 & NGC 5028 & 13 13 45.8 & $-13$ 02 33 & 6433 & 13.72 & & \\
70262 & KUG 2258+193 & 23 01 07.1 & +19 36 33 & 6473 & 15.21 & Sc & \\
3090 & NGC 282 & 00 52 42.1 & +30 38 21 & 6673 & 14.43 & & \\
61167 & NGC 6515 & 17 57 25.2 & +50 43 41 & 6853 & 13.96 & & \\
4808 & AM0118-500 & 01 20 13.4 & $-49$ 49 47 & 7500 & 14.87 & & \\
57841 & CGCG 052-004 & 16 19 48.1 & +05 09 44 & 7494 & 14.76 & & \\
473 & MRK 335 & 00 06 19.5 & +20 12 10 & 7730 & 13.64 & Seyfert 1, S0? & \\
16415 & ESO 033-G003 & 04 57 47.6 & $-73$ 13 54 & 7664 & 14.25 & Possible SA0, behind LMC & \\
55698 & ARK 481 & 15 39 05.1 & +05 34 16 & 7781 & 15.24 & &\\
170383 & & 22 01 23.5  & $-03$ 45 24 & 8001 & 15.13 & Not in NED & \\
7468 & NGC 766 & 01 58 42.0 & +08 20 48 & 8104 & 14.24 & Group member & \\
170381 & & 22 01 05.0 & $-04$ 47 51 & 8248 & 15.46 & Not in NED & \\
65215 & MCG-02-52-019 & 20 41 48.4 & $-13$ 50 49 & 8409 & 14.74 & & \\
7862 & MCG-05-06-002 & 02 03 56.3 & $-31$ 47 09 & 8363 & 14.74 & & \\
60164 & NGC 6363 & 17 22 40.0 & +41 06 06 & 8912 & 15.17 & & \\
57371 & CGCG 320-009 & 16 10 21.0 & +67 50 11 & 8914 & 15.37 & Early spiral? & \\
25560 & CGCG 005-056 & 09 06 39.5 & $-00$ 51 55 & 9058 & 15.37 & NED S0 & \\
54129 & CGCG 165-027 & 15 10 08.8 & +31 53 16 & 9206 & 15.32 & NED S? & \\
&&&&&&&\\
&&&&&&&\\
\tableline

\end{tabular}
\end{center}
\end{table*}

\clearpage


\begin{figure}
\epsscale{0.5}\plotone{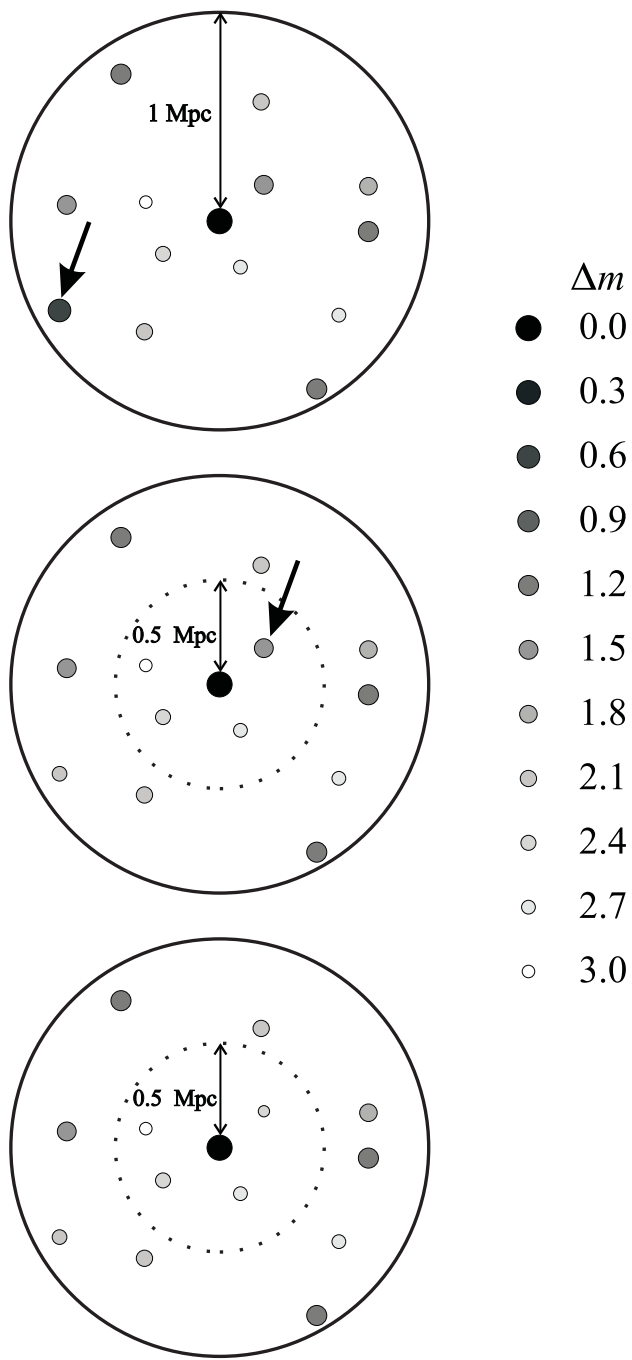}
\caption{The isolation criterion is illustrated in this plot.
The top panel shows the situation when the central
galaxy does not verify the first condition of the criterion
because it has a bright companion (indicated by the arrow), 
with a magnitude difference less than 0.7 within 1 Mpc
of projected separation. In the middle panel, the galaxy
verifies this condition, but fails to be considered
isolated because the second condition is not accomplished: 
it has a companion with a magnitude
difference less than 2.2 within 0.5 Mpc of projected
distance, (again it has been marked with an arrow). 
Finally, the bottom panel shows the case
where the central galaxy verifies the two conditions
of the isolation criterion.}
\label{fig1}
\end{figure}

\begin{figure}
\plotone{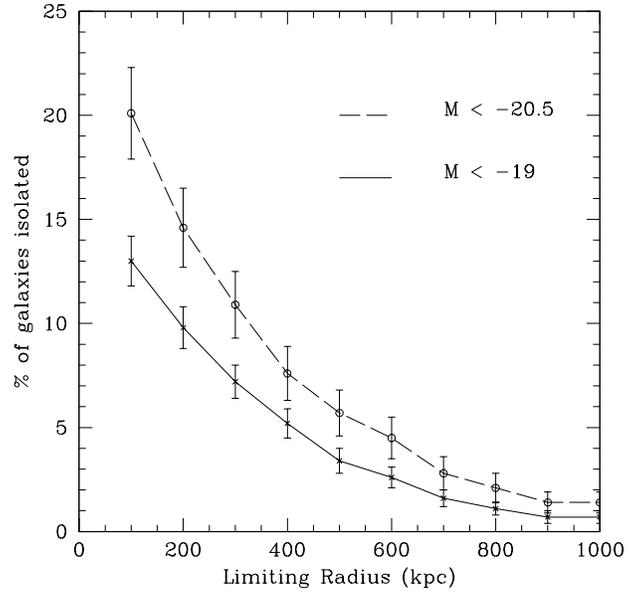}
\caption{Variation of the percentage of elliptical galaxies classified
as isolated as a function of the inner cut-off radius for galaxies within
2.2 magnitudes of the primary. Two values for the limiting absolute
magnitude of the primary are shown.}
\label{fig2}
\end{figure}

\begin{figure}
\plotone{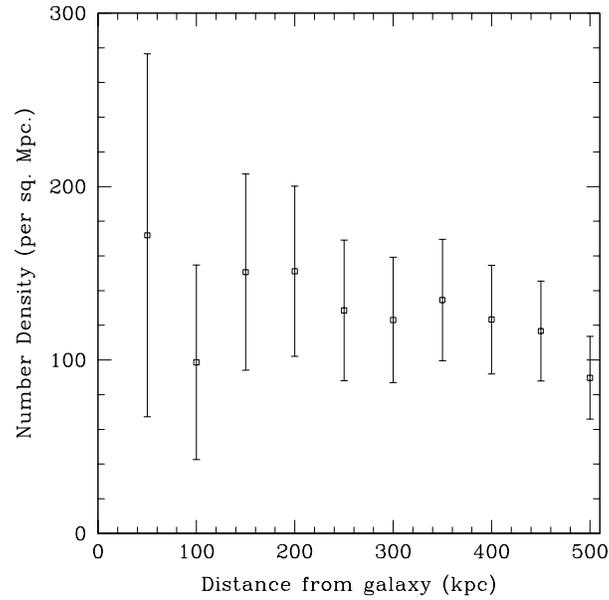}
\caption{Mean radial galaxy density profile for dwarf galaxies surrounding
10 of the 32 galaxies taken to be `isolated'. }
\label{fig3}
\end{figure}

\end{document}